# EAH: A New Encoder based on Adaptive Variable-length Codes

Dragoş Trincă

*Department of Computer Science & Engineering, University of Connecticut Storrs, CT 06269, USA*

**Abstract**

Adaptive variable-length codes associate a variable-length codeword to the symbol being encoded depending on the previous symbols in the input string. This class of codes has been recently presented in [Dragoş Trincă, arXiv:cs.DS/0505007] as a new class of non-standard variable-length codes. New algorithms for data compression, based on adaptive variable-length codes of order one and Huffman's algorithm, have been recently presented in [Dragoş Trincă, ITCC 2004]. In this paper, we extend the work done so far by the following contributions: first, we propose an improved generalization of these algorithms, called EAH$n$. Second, we compute the entropy bounds for EAH$n$, using the well-known bounds for Huffman's algorithm. Third, we discuss implementation details and give reports of experimental results obtained on some well-known corpora. Finally, we describe a parallel version of EAH$n$ using the PRAM model of computation.

## 1 Introduction

With the continuous growth of the Internet, the need of rapid network communications, and the completion of the Human Genome, data compression seems to remain an important and attractive research area.

One of the earliest and most studied data compression techniques is the well-known Huffman's classical algorithm [6]. Even if there are a lot of algorithms which have been developed in the last decades, and which significantly outperform Huffman's algorithm in terms of compression performance, it seems that the classical Huffman coding is *still* the subject of many studies. We give two illustrative examples.

*Email address:* dtrinca@engr.uconn.edu.

First, it is well-known that Huffman coding has been – and still is – successfully used as an intermediate entropy-coding step in the Burrows-Wheeler's compression algorithm introduced in 1994 [4]. After its publication, their algorithm has been improved a lot, and is currently available on multiple platforms [13] as a general-purpose compression algorithm. Second, it seems that Huffman coding is currently used in industry [5] to develop compression algorithms and protocols for rapid satellite communications.

From a practical perspective, using Huffman coding as an intermediate step in other compression schemes seems a bit surprising, since it performs poorly in practice compared with other algorithms. Some of the reasons behind this choice are the following:

(1) Huffman's algorithm has a good running time compared with other compression algorithms. Therefore, when the runtime is a critical parameter in the system, this seems a reasonable choice.
(2) Usually, when used as an intermediate entropy-coding step, it gives good results compared with other compression techniques.

In this paper, we present a new coding technique called EAH$n$ (i.e., Encoder based on Adaptive variable-length codes of order $n$ and Huffman's algorithm), which is a generalization of the algorithms recently presented in [18]. EAH$n$ is *not* intended to be used as a single encoder. Instead, it is intended to replace Huffman coding as an intermediate step in other compression schemes.

The paper is organized as follows. First, we recall some basic definitions and notations related to adaptive variable-length codes (section 2). In section 3, we present in detail the new algorithm EAH$n$, by discussing both offline and online versions. After describing the algorithm in detail, we compute its entropy bounds using the well-known entropy bounds for Huffman's algorithm (section 4). Implementation details and reports of experimental results obtained on some well-known corpora are provided in sections 5 and 6. As we shall see, using EAH$n$ instead of Huffman's algorithm, either as an intermediate step in other compression schemes or as a single encoder, is preferred, since it gives significantly better results. In section 7, we describe a parallel version of EAH$n$ using the PRAM model of computation. Finally, in the last section, we discuss some future work directions.

## 2 Adaptive variable-length codes

Adaptive variable-length codes have been recently presented in [18] as a new class of non-standard variable-length codes. The aim of this section is to briefly review some basic definitions and notations. For more details, the reader is referred to [17,18].

We denote by $|S|$ the *cardinality* of the set $S$; if $x$ is a string of finite length, then $|x|$ denotes the length of $x$. The *empty string* is denoted by $\lambda$. For an alphabet $\Sigma$, we denote by $\Sigma^n$ the set $\{s_1 s_2 \ldots s_n \mid s_i \in \Sigma \text{ for all } i\}$, by $\Sigma^*$ the set $\bigcup_{n=0}^{\infty} \Sigma^n$, and by $\Sigma^+$ the set $\bigcup_{n=1}^{\infty} \Sigma^n$, where $\Sigma^0$ denotes the set $\{\lambda\}$. Also, we denote by $\Sigma^{\leq n}$ the set $\bigcup_{i=0}^{n} \Sigma^i$, and by $\Sigma^{\geq n}$ the set $\bigcup_{i=n}^{\infty} \Sigma^i$.

Let $X$ be a finite and nonempty subset of $\Sigma^+$, and $w \in \Sigma^+$. A *decomposition of $w$ over $X$* is any sequence of strings $u_1, u_2, \ldots, u_h$ with $u_i \in X$ for all $i$, such that $w = u_1 u_2 \ldots u_h$. A *code over $\Sigma$* is any nonempty set $C \subseteq \Sigma^+$ such that each string $w \in \Sigma^+$ has at most one decomposition over $C$. A *prefix code over $\Sigma$* is any code $C$ over $\Sigma$ such that no string in $C$ is a proper prefix of another string in $C$. If $u, v$ are two strings, then we denote by $u \cdot v$, or simply by $uv$ the concatenation of $u$ with $v$.

**Definition 1** *Let $\Sigma$ and $\Delta$ be two alphabets. A function $c : \Sigma \times \Sigma^{\leq n} \to \Delta^+$, with $n \geq 1$, is called an* adaptive variable-length code of order $n$ *if its extension $\overline{c} : \Sigma^* \to \Delta^*$, given by*

- $\overline{c}(\lambda) = \lambda$,
- $\overline{c}(\sigma_1 \ldots \sigma_m) = c(\sigma_1, \lambda) c(\sigma_2, \sigma_1) \ldots c(\sigma_{n+1}, \sigma_1 \ldots \sigma_n) \ldots c(\sigma_m, \sigma_{m-n} \ldots \sigma_{m-1})$,

*for all strings $\sigma_1 \ldots \sigma_m \in \Sigma^+$, is injective.*

As it is clearly specified in the definition above, an adaptive code of order $n$ associates a variable-length codeword to the symbol being encoded depending on the previous $n$ symbols in the input data string. Let us now give an example in order to better understand this mechanism.

**Example 2** *Let $\Sigma = \{\mathtt{a}, \mathtt{b}, \mathtt{c}\}$, $\Delta = \{0, 1\}$ be alphabets, and consider the function $c : \Sigma \times \Sigma^{\leq 1} \to \Delta^+$ given by Table 1.*

Table 1
An adaptive variable-length code of order one

| $\Sigma \backslash \Sigma^{\leq 1}$ | a   | b  | c   | $\lambda$ |
|---|---|---|---|---|
| a | 010 | 10 | 0   | 11 |
| b | 011 | 01 | 100 | 01 |
| c | 11  | 11 | 101 | 00 |

*One can verify that $\overline{c}$ is injective, and according to Def. 1, $c$ is an adaptive variable-length code of order one. Let $x = \mathtt{abbaba} \in \Sigma^+$ be an input data string. Using Def. 1, we encode the string $x$ by*

$$\overline{c}(x) = c(\mathtt{a}, \lambda) c(\mathtt{b}, \mathtt{a}) c(\mathtt{b}, \mathtt{b}) c(\mathtt{a}, \mathtt{b}) c(\mathtt{b}, \mathtt{a}) c(\mathtt{a}, \mathtt{b}) = 11011011001110.$$

**Example 3** *Let $\Sigma = \{{\tt a}, {\tt b}\}$, $\Delta = \{0, 1\}$ be alphabets, and $c : \Sigma \times \Sigma^{\leq 2} \to \Delta^+$ a function given by Table 2. It is easy to verify that $\overline{c}$ is injective, and according to Def. 1, c is an adaptive variable-length code of order two.*

Table 2
An adaptive variable-length code of order two

| $\Sigma \backslash \Sigma^{\leq 2}$ | a | b | aa | ab | ba | bb | $\lambda$ |
|---|---|---|---|---|---|---|---|
| a | 010 | 100 | 0 | 11 | 110 | 00 | 001 |
| b | 011 | 010 | 10 | 01 | 010 | 10 | 111 |

*Let $x = {\tt ababba} \in \Sigma^+$ be an input data string. Using Def. 1, we encode the string $x$ by*

$$\overline{c}(x) = c({\tt a}, \lambda)c({\tt b}, {\tt a})c({\tt a}, {\tt ab})c({\tt b}, {\tt ba})c({\tt b}, {\tt ab})c({\tt a}, {\tt bb}) = 001011110100100.$$

Let $c : \Sigma \times \Sigma^{\leq n} \to \Delta^+$ be an adaptive variable-length code of order $n$, with $n \geq 1$. We denote by $C_{c,\sigma_1\sigma_2\ldots\sigma_h}$ the set $\{c(\sigma, \sigma_1\sigma_2\ldots\sigma_h) \mid \sigma \in \Sigma\}$, for all $\sigma_1\sigma_2\ldots\sigma_h \in \Sigma^{\leq n} - \{\lambda\}$, and by $C_{c,\lambda}$ the set $\{c(\sigma, \lambda) \mid \sigma \in \Sigma\}$. We write $C_{\sigma_1\sigma_2\ldots\sigma_h}$ instead of $C_{c,\sigma_1\sigma_2\ldots\sigma_h}$, and $C_\lambda$ instead of $C_{c,\lambda}$ whenever there is no confusion. Also, let us denote by $AC(\Sigma, \Delta, n)$ the set of all adaptive variable-length codes of order $n$ from $\Sigma$ to $\Delta$. The proof of the following important theorem can be found in [17].

**Theorem 4** *Let $\Sigma$ and $\Delta$ be two alphabets, and let $c : \Sigma \times \Sigma^{\leq n} \to \Delta^+$ be a function, $n \geq 1$. If $C_u$ is prefix code, for all $u \in \Sigma^{\leq n}$, then $c \in AC(\Sigma, \Delta, n)$.*

## 3 EAH$n$: an encoder based on adaptive variable-length codes

As we have already pointed out, the aim of this section is to present a new lossless data compression algorithm, called EAH$n$, which is actually a generalization, from adaptive variable-length codes of order one to adaptive variable-length codes of any order, of the algorithms presented in [18].

Let us first fix some very useful notation, which will be used in the description of the algorithms. Let $\mathcal{U} = (u_1, u_2, \ldots, u_k)$ be a $k$-tuple. We denote by $\mathcal{U}.i$ the $i$-th component of $\mathcal{U}$, that is, $\mathcal{U}.i = u_i$ for all $i \in \{1, 2, \ldots, k\}$. The 0-tuple is denoted by (). The length of a tuple $\mathcal{U}$ is denoted by $Len(\mathcal{U})$. If q is a component or a tuple, and $i \in \{1, 2, \ldots, k\}$, then we define $\mathcal{U} \triangleleft q$ and $\mathcal{U} \triangleright i$ by

- $\mathcal{U} \triangleleft q = (u_1, \ldots, u_k, q)$,
- $\mathcal{U} \triangleright i = (u_1, \ldots, u_{i-1}, u_{i+1}, \ldots, u_k)$.

If $\mathcal{A}$ denotes an algorithm and $x$ its input, then we denote by $\mathcal{A}(x)$ its output. Also, $\mathbb{N}$ denotes the set of natural numbers.

**Algorithm Huffman.** *It is well-known that Huffman's algorithm takes as input a tuple $\mathcal{F} = (f_1, f_2, \ldots, f_n)$ of frequencies, and returns a tuple $\mathcal{V} = (v_1, v_2, \ldots, v_n)$ of codewords, such that $v_i$ is the codeword corresponding to the symbol with the frequency $f_i$, for all $i \in \{1, 2, \ldots, n\}$. This procedure will be used by our encoder EAHn in order to construct an adaptive variable-length code of order $n$. More precisely, if $c : \Sigma \times \Sigma^{\leq n} \to \Delta^+$ denotes the code constructed by EAHn, then we apply the Huffman's algorithm to each of the sets $\{Freq(u\sigma) \geq 1 \mid \sigma \in \Sigma\}$, where $u \in \Sigma^n$ and $Freq(u\sigma)$ denotes the number of occurrences of $u\sigma$ in the input string.*

**Algorithm EAHn.** *Let $\Sigma = \{\sigma_1, \sigma_2, \ldots, \sigma_p\}$ be an alphabet, and let $x$ be a string over $\Sigma$. Let $q$ be the number of symbols occurring in $x$ (thus, $q \leq p$). Let us explain the main idea of our scheme. Consider that $u \in \Sigma^n$ is some substring of the input string $x$. Also, let us denote by $Follow(u)$ the set of symbols that follow the substring $u$ in $x$. For each symbol $c \in Follow(u)$, let us denote by $Freq(uc)$ the frequency of the substring $uc$ in $x$. One can easily remark that $Follow(u)$ cannot contain more than $q$ symbols. Moreover, in most cases, the number of symbols in $Follow(u)$ is significantly smaller than $q$. Instead of applying the Huffman's algorithm to the frequencies of the $q$ symbols occurring in $x$, we apply it to each of the sets $\{Freq(uc) \mid c \in Follow(u)\}$, since every such set usually has a smaller number of elements. If $code(c, u)$ is the codeword associated to the substring $uc$, then we encode $c$ by $code(c, u)$ whenever it is preceded by $u$. Therefore, in most cases, we get smaller codewords.*

*Thus, Huffman's algorithm is actually applied to every substring $u$ of length $n$ occurring in $x$. We associate to each symbol a set of codewords, and encode each symbol with one of the codewords in its set, depending on the previous $n$ symbols occurring in the input string.*

*The complete algorithm is given in Fig. 1. Let us now explain what exactly the algorithm performs at each step. The first three steps are aimed to initialize the functions needed. Note that the function $d$ actually allows us to access the elements of $\Sigma^n$ in a certain order. In the fourth step, $b(x_i, x_{i-n} \ldots x_{i-1})$ is switched to 1, since the substring $x_{i-n} \ldots x_{i-1} x_i$ occurs at least once in $x$, and the frequency of $x_{i-n} \ldots x_{i-1} x_i$ is incremented. In the fifth step, for every substring $d(j)$ occurring in $x$, we apply the Huffman's algorithm to the set $\{Freq(d(j)\sigma) \mid \sigma \in Follow(d(j))\}$. In the next two steps, $\mathcal{Y}$ is a tuple of codewords constructed as follows. If $c \in \Sigma$ and $u \in \Sigma^n$, then $a(c, u)$ is appended to $\mathcal{Y}$ if and only if $a(c, u) \neq \lambda$, that is, if $c \in Follow(u)$ and $|Follow(u)| \geq 2$. Finally, in the last step, $Z$ denotes the compression of $x_{n+1} \ldots x_t$.*

*So, the compression of the string $x$ is actually $Z$. The first three components*

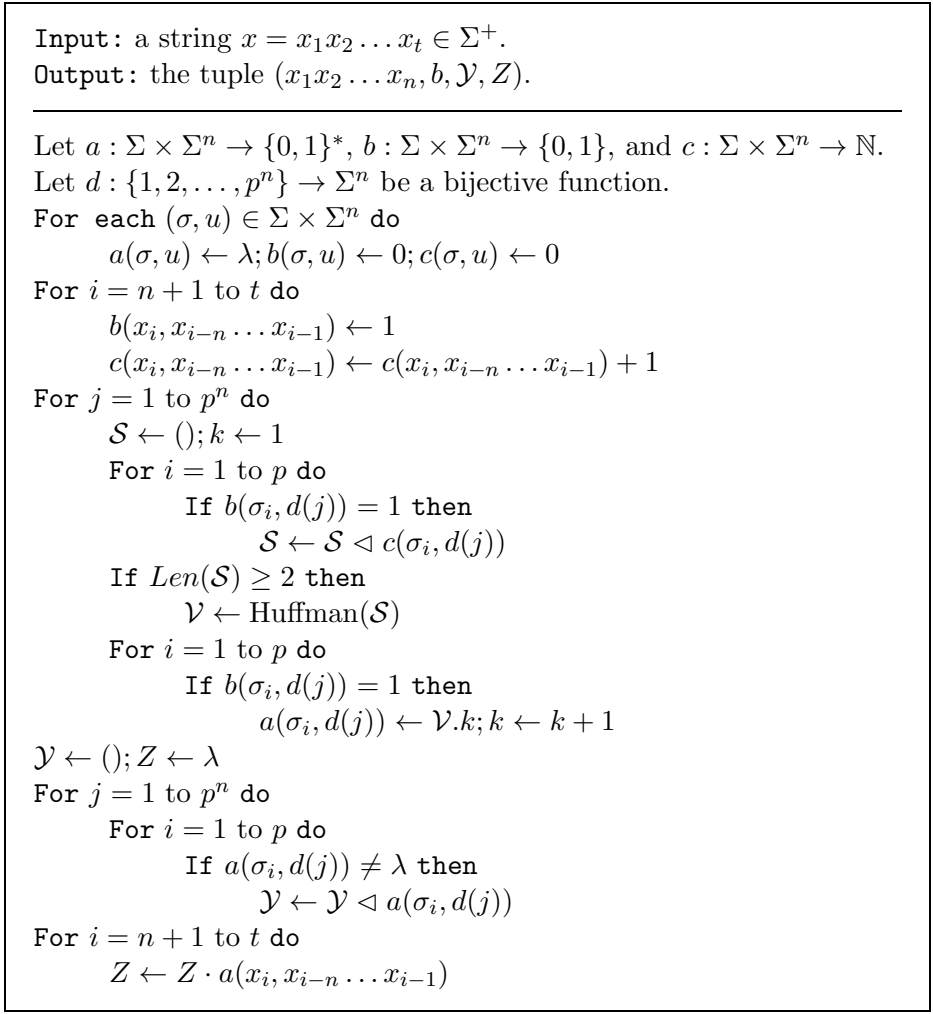

Fig. 1. EAH$n$

of the output (i.e., $x_1 x_2 \ldots x_n$, $b$, and $\mathcal{Y}$) are only needed when decoding $Z$ into $x$. The decoding procedure is obtained by following the same steps in reverse order.

Let us now take an example in order to better understand the description above.

**Example 5** *Let $\Sigma = \{\mathtt{a}, \mathtt{b}\}$ be an alphabet, $x = \mathtt{baabbabab} \in \Sigma^+$ an input data string, and let us take $n = 2$. After applying EAH2 to the string $x$, we get the results reported in Tables 3, 4, and 5.*

Table 3
The function $a$ after applying EAH2 to $x$

| $\Sigma \backslash \Sigma^2$ | aa | ab | ba | bb |
|---|---|---|---|---|
| a | $\lambda$ | 0 | 0 | $\lambda$ |
| b | $\lambda$ | 1 | 1 | $\lambda$ |

Table 4
The function $b$ after applying EAH2 to $x$

| $\Sigma\backslash\Sigma^2$ | aa | ab | ba | bb |
|---|---|---|---|---|
| a | 0 | 1 | 1 | 1 |
| b | 1 | 1 | 1 | 0 |

Table 5
The function $c$ after applying EAH2 to $x$

| $\Sigma\backslash\Sigma^2$ | aa | ab | ba | bb |
|---|---|---|---|---|
| a | 0 | 1 | 1 | 1 |
| b | 1 | 1 | 2 | 0 |

*Let us now explain these results by considering the third column of each table, i.e., the column corresponding to the substring* ba. *In Table 4, $b(\mathtt{a},\mathtt{ba}) = 1$ and $b(\mathtt{b},\mathtt{ba}) = 1$, since the substrings* baa *and* bab *both occur at least once in $x$. In Table 5, $c(\mathtt{a},\mathtt{ba}) = 1$ is the frequency of* baa *in $x$, and $c(\mathtt{b},\mathtt{ba}) = 2$, since* bab *occurs twice in $x$. Thus, applying the Huffman's algorithm to the set of frequencies $\{1,2\}$, we encode the symbol* a *by $a(\mathtt{a},\mathtt{ba}) = 0$ whenever it is preceded by* ba. *Also, we encode the symbol* b *by $a(\mathtt{b},\mathtt{ba}) = 1$ whenever it is preceded by* ba.

*Considering that the function $d$ is given by $d(1) = \mathtt{aa}$, $d(2) = \mathtt{ab}$, $d(3) = \mathtt{ba}$, and $d(4) = \mathtt{bb}$, one can verify that the output of* EAH2 *in this example is the 4-tuple*

$$(\mathtt{ba}, b, (0,1,0,1), 01101),$$

*where $b$ is the function given above. Also, one can remark that the function $b$ can be encoded using $p^{n+1}$ bits. In our example, $b$ can be encoded by $2^3 = 8$ bits, since $p = 2$ and $n = 2$.*

**Remark 6** *Let $x = x_1 x_2 \ldots x_t$. Given that the size $p$ of the alphabet and the parameter $n$ do not depend on the input, we can conclude that the runtime of EAHn is linear in the size of the input. More precisely, for a fixed value of $n$, the runtime of EAHn is $O(t)$. If we consider the the input consists of both $n$ and $t$, then the runtime is $O(max\{p^n, t\})$. Thus, given that $p$ is a constant, we conclude that the runtime is either exponential in the size of $n$ or linear in the size of the input. However, in practice $p$ and $n$ are very small numbers, so it is unlikely that we need more than $O(t)$ time.*

Even though EAHn is based on adaptive variable-length codes, i.e., it encodes the current symbol adaptively by taking into account only the last few sym-bols, it is an *offline* algorithm, since the codewords are chosen only after the input string has been already processed. Let us explain this in detail. Let $c$ be the current symbol and $u \in \Sigma^n$ the string of length $n$ preceding $c$. Then, the codeword associated to $c$ is chosen based on the frequency of $uc$ in the entire input. EAH$n$ can be easily transformed into an *online* (i.e., dynamic) algorithm as follows: we associate a codeword to the current symbol $c$ based *only* on the frequency of $uc$ in the text already processed. More precisely, if $x = x_1 x_2 \ldots x_t$ is the input string and $x_i$ is the current symbol, then the codeword associated to $x_i$ is chosen based on the frequency of $x_{i-n} \ldots x_{i-1} x_i$ in $x_1 x_2 \ldots x_i$.

It is well-known [19] that, except for short messages, the online Huffman coding always produces a longer encoding than the offline version, the reason being that the offline version is optimal in this respect. In practice, the online version of EAH$n$ is a good choice whenever the runtime is a critical parameter in the system. Moreover, as described in [19], an online compressor can be used to encode the message in *real time*. Thus, whenever the runtime is one of the most critical parameters in the system, the online version of EAH$n$ is preferred over the static one, since the message can be dynamically encoded.

## 4 Entropy bounds for EAH$n$

This section focuses on computing the entropy bounds for EAH$n$. Since the EAH$n$ encoder is based on Huffman's algorithm, let us first recall the well-known bounds for Huffman's algorithm.

**Definition 7** *Let $\Sigma$ be an alphabet, $x$ an input data string of length $t$ over $\Sigma$, and $k$ the length of the encoder output. The* compression rate *of $x$, denoted by $R(x)$, is defined by*

$$R(x) = \frac{k}{t}. \tag{1}$$

Let $R_\text{H}(x)$ be the compression rate in codebits per datasample, computed after encoding the string $x$ by Huffman's algorithm. One can obtain upper and lower bounds on $R_\text{H}(x)$ before encoding the data string $x$, by computing the entropy denoted by $E_\text{H}(x)$. Let us consider that $x$ is a string of length $t$, $(F_1, F_2, \ldots, F_h)$ is the $h$-tuple of frequencies corresponding to the symbols in $x$, and $k$ is the length of the Huffman's encoder output. The entropy $E_\text{H}(x)$ of $x$ is defined by

$$E_\text{H}(x) = \frac{1}{t} \sum_{i=1}^{h} F_i \log_2(\frac{t}{F_i}). \tag{2}$$

Let $L_i$ be the length of the codeword associated by Huffman's algorithm to the symbol with the frequency $F_i$, for all $i \in \{1, 2, \ldots, h\}$. Then, the compression rate $R_H(x)$ can be re-written by

$$R_H(x) = \frac{1}{t} \sum_{i=1}^{h} F_i L_i. \tag{3}$$

Relating the entropy $E_H(x)$ to the compression rate $R_H(x)$, we obtain the inequalities

$$E_H(x) \leq R_H(x) \leq E_H(x) + 1. \tag{4}$$

Let $\Sigma = \{\sigma_1, \sigma_2, \ldots, \sigma_p\}$ be the same alphabet of size $p$, and let $x = x_1 x_2 \ldots x_t$ be a string over $\Sigma$. Let us denote by $R_{EAHn}(x)$ the compression rate obtained after encoding $x$ by EAH$n$. More precisely, $R_{EAHn}(x)$ is given by

$$R_{EAHn}(x) = \frac{Z}{t}, \tag{5}$$

where $Z$ denotes the fourth component of the EAH$n$ output, i.e., the encoding of $x_{n+1} x_{n+2} \ldots x_t$. Also, let us denote by $E_{EAHn}(x)$ the EAH$n$ entropy of $x$.

We can obtain upper and lower bounds on $R_{EAHn}(x)$ before encoding $x$, by computing the entropy $E_{EAHn}(x)$. Given that EAH$n$ uses Huffman's algorithm to associate a set of codewords to each substring of length $n$ occurring in $x_1 x_2 \ldots x_{t-1}$, it is clear that the entropy $E_{EAHn}(x)$ can be computed as a sum of entropies. Let us explain this in detail, since it will help us to derive the final formula for $E_{EAHn}(x)$.

Let $J : \{0, 1, \ldots, p^n - 1\} \to \Sigma^n$ be a fixed bijective function. In other words, $J(i)$ identifies a certain substring of length $n$. Thus, each substring $J(i)$ of length $n$ can be uniquely identified by its index $i$. Let $A$ be the set of those indexes $i$ such that $J(i)$ occurs at least once in $x_1 x_2 \ldots x_{t-1}$. It is now clear that

$$E_{EAHn}(x) = \sum_{i \in A} E_{EAHn}(x, i), \tag{6}$$

where $E_{EAHn}(x, i)$ denotes the entropy corresponding to those positions in $x$ preceded by $J(i)$. Now, in order to establish the formula for $E_{EAHn}(x)$, we only need to compute $E_{EAHn}(x, i)$. Let $C(i)$ denote the set of those $j$'s such that $J(i) \cdot \sigma_j$ occurs at least once in $x$. In other words, $C(i)$ denotes the set of those symbols in $x$ preceded by $J(i)$. The cardinality of $C(i)$ corresponds to the number $h$ in the general case described above. Also, let $N(i)$ be the total number of positions in $x$ preceded by $J(i)$. $N(i)$ corresponds to the number $t$ in the general case. Finally, if we denote by $F(i,j)$ the frequency of $J(i) \cdot \sigma_j$ in $x$, we can conclude that

$$E_{\text{EAH}n}(x, i) = \frac{\sum_{j \in C(i)}[F(i,j) \log_2 \frac{N(i)}{F(i,j)}]}{N(i)}. \tag{7}$$

$F(i,j)$ corresponds to the number $F(i)$ in the general case. Thus, the entropy $E_{\text{EAH}n}(x)$ is actually a sum of entropies, where each such local entropy is computed similarly to the general case.

## 5 Implementation details for EAH1

In this section, we give complete details regarding the structure of the files compressed with EAH1. Specifically, the structure looks as shown in Table 6. Thus, we describe precisely how the output of EAH1 is encoded in a file.

Table 6
The structure of the files compressed with EAH1

| Z1 | Z2 | Z3 | Z4 | Z5 | Z6 | Z7 | Z8 | Z9 |
|----|----|----|----|-----|-----|----|----|----|
| 8  | NF | 8  | 8  | 256 | 256 | NF | NF | NF |

In our implementation, the size of the alphabet is 256, so each symbol requires 8 bits. In the table above, each of the fields denoted Z1,Z2,…,Z9 has associated its size in bits. NF (not fixed) denotes the size of the fields whose length depends on the input. Let us now describe each field separately.

**Z1.** This field has a fixed length (8 bits), and specifies how many padding bits are used in the field Z2. Thus, the number encoded by Z1 gives us the length of Z2.

**Z2.** This second field consists of a sequence of padding bits, which are appended to the whole structure so that the total number of bits in Z1,…,Z9 is a multiple of 8. So, the length of Z2 is a number between 0 and 7.

**Z3.** If the input data string for EAH1 is $x = x_1 x_2 \ldots x_t$, then this field encodes the symbol $x_1$, that is, the first component of the output.

**Z4.** The fourth field gives us the maximum length of a codeword in $\mathcal{Y}$, where $\mathcal{Y}$ is the third component of the output. Let us denote by $MAXLC$ this number.

**Z5.** For each symbol in the alphabet, we need a bit in order to specify if it occurs at least once in $x_1 x_2 \ldots x_{t-1}$. Let us denote by $NC$ the number of bits 1 in this field.

**Z6.** For each symbol in the alphabet, we need a bit in order to specify if it occurs at least once in $x_2 x_3 \ldots x_t$. Let us denote by $NL$ the number of bits 1 in this field.

**Z7.** This field consists of $NL * NC$ bits, and together with Z5 and Z6 encodes the function $b$, that is, the second component of the output.

**Z8.** The length of this field is $Len(\mathcal{Y}) * (MAXLC + 1)$, since each of the codewords in $\mathcal{Y}$ is encoded by $MAXLC+1$ bits. Actually, $Len(\mathcal{Y})$ is at most the number of bits 1 in Z7. Let us now explain how exactly a codeword in $\mathcal{Y}$ is encoded using only $MAXLC+1$ bits. If $cw = B_1 B_2 \ldots B_i$ is a codeword in $\mathcal{Y}$, then we know that $i \leq MAXLC$. Let us denote by $\overline{B}$ the complement of the bit $B$. If $i = MAXLC$, then we encode $cw$ by $B_1 B_1 B_2 \ldots B_i$. Otherwise, if $i < MAXLC$, we encode $cw$ by $B_1 u B_1 B_2 \ldots B_i$, where $u = \overline{B_1} \ldots \overline{B_1}$ is a sequence of length $MAXLC - i$. The decoding works as follows. Suppose that $C_1 C_2 \ldots C_{MAXLC+1}$ is a sequence of bits denoting the encoding of a codeword $cw$, and let $j$ be such that $C_2 = \ldots = C_j$, but $C_j \neq C_{j+1}$ (if there does not exist such an index, we know that the codeword $cw$ is $C_2 \ldots C_{MAXLC+1}$). If $C_1 = C_2$, then $cw$ is $C_2 \ldots C_{MAXLC+1}$. Otherwise, if $C_1 \neq C_2$, we know that $cw = C_{j+1} \ldots C_{MAXLC+1}$.

**Z9.** Finally, the last field denotes the compression of $x_2 x_3 \ldots x_t$. Precisely, this field is actually $Z$, the last component of the output.

## 6  Experimental results

Given that EAH$n$ is an offline algorithm, it seems natural to compare it with Huffman's classical algorithm. For experiments, we have chosen two of the most known corpora: The Calgary Compression Corpus (CCC) and The Large Canterbury Corpus (LCC) [20]. All the comparisons provided in the Tables 7, 8, and 9 show specific differences between EAH1 and Huffman's classical encoder. As one can see, the EAH1 encoder gives significantly better results on both corpora. The improvement is approximately 21.20% for CCC, and 18.99% for LCC.

Table 7
Results of compressing three files of the Large Canterbury Corpus (LCC)

| File | Size (bytes) | HUFFMAN | EAH1 | Improvement (%) |
|---|---|---|---|---|
| E.coli | 4,638,690 | 1,159,677 | 1,159,748 | — |
| bible.txt | 4,047,392 | 2,218,595 | 1,690,454 | 23.80 |
| world192.txt | 2,473,400 | 1,558,845 | 1,148,918 | 26.29 |
| Total | 11,159,482 | 4,937,117 | 3,999,120 | — |

Table 8
Results of compressing fourteen ASCII files of the Calgary Compression Corpus (CCC)

| | Size | | | Improvement |
|---|---|---|---|---|
| File | (bytes) | HUFFMAN | EAH1 | (%) |
| bib | 111,261 | 72,936 | 49,540 | 32.07 |
| book1 | 768,771 | 438,592 | 351,144 | 19.93 |
| book2 | 610,856 | 368,507 | 294,717 | 20.02 |
| news | 377,109 | 246,580 | 200,372 | 18.73 |
| paper1 | 53,161 | 33,530 | 27,042 | 19.34 |
| paper2 | 82,199 | 47,812 | 38,511 | 19.45 |
| paper3 | 46,526 | 27,435 | 22,481 | 18.05 |
| paper4 | 13,286 | 8,003 | 7,584 | 5.23 |
| paper5 | 11,954 | 7,593 | 7,212 | 5.01 |
| paper6 | 38,105 | 24,212 | 20,164 | 16.71 |
| progc | 39,611 | 26,090 | 19,865 | 23.85 |
| progl | 71,646 | 43,148 | 31,408 | 27.20 |
| progp | 49,379 | 30,395 | 21,740 | 28.47 |
| trans | 93,695 | 65,431 | 43,055 | 34.19 |
| Total | 2,367,559 | 1,440,264 | 1,134,835 | — |

Table 9
Improvements per corpus

| Corpus | total compressed size HUFFMAN | total compressed size EAH1 | Improvement (%) |
|---|---|---|---|
| CCC | 1,440,264 | 1,134,835 | 21.20 |
| LCC | 4,937,117 | 3,999,120 | 18.99 |
| Total | 6,310,138 | 5,132,328 | — |

## 7 An EREW PRAM version of EAH$n$

This section focuses on describing a parallel version of EAH$n$, using the well-known PRAM model of computation [7]. As we shall see, EAH$n$ is highly parallelizable, mostly because during its execution, Huffman's algorithm is applied to *disjoint* sets of frequencies.

Let us first recall some basic concepts. The *shared-memory model* consists of a number of processors that have access to a single shared memory unit, usually referred to as *global memory*. Each processor has its own *local memory* and can execute its own local program. Processors communicate by exchanging data through the shared memory unit. Each processor is identified by its unique *index*, which is available in its local memory. A general view of this model is given in Fig. 2.

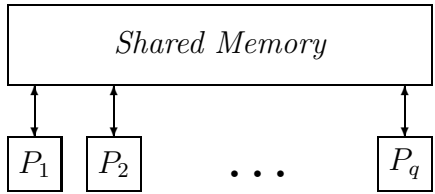

Fig. 2. The Shared-memory Model

If all the processors work synchronously under the control of a common clock, then the shared memory model is usually called the *parallel random-access machine* (PRAM) model. Throughout this section, we use only the *exclusive-read exclusive-write* (EREW) PRAM submodel, i.e., the PRAM submodel that does not allow any simultaneous access to a single memory location. For more details, the reader is referred to [7].

Let $\Sigma = \{\sigma_1, \sigma_2, \ldots, \sigma_p\}$ be the same alphabet of size $p$, and $x = x_1 x_2 \ldots x_t$ an input string over $\Sigma$. Suppose that we have $q = p^n$ EREW PRAM processors, denoted $p_1, p_2, \ldots, p_q$. Then, we can assign one processor to each substring $u \in \Sigma^n$, since the cardinality of $\Sigma^n$ is $p^n$. Let $d : \{1, 2, \ldots, p^n\} \to \Sigma^n$ be a fixed bijective function. For each $i \in \{1, 2, \ldots, p^n\}$, processor $p_i$ is assigned to the string $d(i)$.

For each $u \in \Sigma^n$ that occurs at least once in $x_1 x_2 \ldots x_{t-1}$, EAH$n$ applies the Huffman's algorithm to the set of frequencies $\{Freq(uc) \mid c \in Follow(u)\}$, as described in section 3. Therefore, in the parallel version, processor $p_i$ applies the Huffman's algorithm to the set $\{Freq(d(i)c) \mid c \in Follow(d(i))\}$ if $d(i)$ occurs at least once in $x_1 x_2 \ldots x_{t-1}$.

The complete algorithm is given in Fig. 3. Note that the functions $a$, $b$, $c$, and $d$ must reside in the global memory, such that each processor can access its corresponding memory locations. Given that the processors do not read from (or write into) the same memory locations, we conclude that our algorithm runs under the EREW PRAM model.

The first two steps are aimed to allocate the necessary space in the shared memory for the functions $a$, $b$, $c$, and $d$. The third step is a parallel one. More precisely, each processor $p_i$ initializes $a(\Sigma, d(i))$, $b(\Sigma, d(i))$, and $c(\Sigma, d(i))$. As one can remark, the fourth step is not executed in parallel. Therefore, we can consider that it is executed by processor $p_1$.

---

Input: a string $x = x_1 x_2 \ldots x_t \in \Sigma^+$.
Output: the tuple $(x_1 x_2 \ldots x_n, b, \mathcal{Y}, Z)$.

---

Let $a : \Sigma \times \Sigma^n \to \{0,1\}^*$, $b : \Sigma \times \Sigma^n \to \{0,1\}$, and $c : \Sigma \times \Sigma^n \to \mathbb{N}$ be three functions with the necessary space allocated in the global memory.
Let $d : \{1, 2, \ldots, p^n\} \to \Sigma^n$ be a bijective function with necessary space allocated in the global memory.
For each $u \in \Sigma^n$ pardo
    For each $\sigma \in \Sigma$ do
        $a(\sigma, u) \leftarrow \lambda; b(\sigma, u) \leftarrow 0; c(\sigma, u) \leftarrow 0$
For $i = n + 1$ to $t$ do
    $b(x_i, x_{i-n} \ldots x_{i-1}) \leftarrow 1$
    $c(x_i, x_{i-n} \ldots x_{i-1}) \leftarrow c(x_i, x_{i-n} \ldots x_{i-1}) + 1$
For $j = 1$ to $p^n$ pardo
    $\mathcal{S} \leftarrow (); k \leftarrow 1$
    For $i = 1$ to $p$ do
        If $b(\sigma_i, d(j)) = 1$ then
            $\mathcal{S} \leftarrow \mathcal{S} \triangleleft c(\sigma_i, d(j))$
    If $Len(\mathcal{S}) \geq 2$ then
        $\mathcal{V} \leftarrow \text{Huffman}(\mathcal{S})$
    For $i = 1$ to $p$ do
        If $b(\sigma_i, d(j)) = 1$ then
            $a(\sigma_i, d(j)) \leftarrow \mathcal{V}.k; k \leftarrow k + 1$
$\mathcal{Y} \leftarrow (); Z \leftarrow \lambda$
For $j = 1$ to $p^n$ pardo
    $\mathcal{Y}_j \leftarrow ()$
    For $i = 1$ to $p$ do
        If $a(\sigma_i, d(j)) \neq \lambda$ then
            $\mathcal{Y}_j \leftarrow \mathcal{Y}_j \triangleleft a(\sigma_i, d(j))$
For $j = 1$ to $p^n$ do
    Append the components of $\mathcal{Y}_j$ to $\mathcal{Y}$
For $i = n + 1$ to $t$ do
    $Z \leftarrow Z \cdot a(x_i, x_{i-n} \ldots x_{i-1})$

Fig. 3. An EREW PRAM version of EAH$n$

In the fifth step, each processor $p_i$ applies the Huffman's algorithm to the set $\{Freq(d(i)c) \mid c \in Follow(d(i))\}$ if and only if this set has at least two elements. The last parallel step is step 7, which constructs the output tuple $\mathcal{Y}$. Steps 6, 8, and 9 are executed only by one of the processors, i.e., they are not parallel steps. For example, step 8 is not executed in parallel since the components of $\mathcal{Y}$ must be in a certain order.

**Runtime.** It is clear that the runtime is $O(max\{p^n, t\})$, where $p$ is the constant size of the alphabet. However, given that most of the steps are executed in parallel, the constant hidden under the $O$-notation is significantly smaller than in the sequential version of EAH$n$. For example, in the third step, Huffman's algorithm is applied in parallel to different sets of frequencies. Given that this is one of the most time-consuming steps, it is clear that it will reduce significantly the runtime.

## 8 Conclusions and future work

Adaptive variable-length codes have been recently presented in [17,18] as a new class of non-standard variable-length codes. New algorithms for data compression, based on adaptive codes of order one and Huffman's algorithm, have been also presented in [18]. In this paper, we extended the work done so far by the following contributions: first, we proposed an improved generalization of the algorithms presented in [18], called EAH$n$. Second, we computed the entropy bounds for EAH$n$, using the well-known bounds for Huffman's algorithm. Third, we discussed implementation details and gave reports of experimental results obtained on some well-known corpora. Finally, we described a parallel version of EAH$n$ using the PRAM model of computation.

One of the most ambitious future plans is to investigate the possibility of incorporating the algorithms presented here into the projects currently developed by some industrial laboratories [5].